\documentstyle[prl,multicol,aps,epsf]{revtex}
\begin{document}
\title{Spin Ordering and Quasiparticles in
Spin Triplet Superconducting Liquids}
\author{Fei Zhou }
\address{ITP, Minnaert building, Leuvenlaan 4,
3584 CE Utrecht, The
Netherlands}
\date{\today}
\maketitle
\begin{abstract}

Spin ordering and its effect on low energy quasiparticles 
in a p-wave superconducting liquid
are investigated.
We study the properties of a novel $2D$ p-wave superconducting
liquid where the ground state is spin rotation invariant. 
In quantum spin disordered liquids,
the low energy quasiparticles are bound states of the bare Bogolubov- De 
Gennes (${\em BdeG}$)
quasiparticles and zero energy skyrmions, which are charge neutral
bosons at the low energy limit.
Further more, spin collective excitations are fractionalized ones
carrying a half spin and obeying fermionic statistics.
In thermally spin disordered limits, the quasi-particles are 
bound states of bare ${\em BdeG}$ quasi-particles.
The latter situation can be realized in some layered p-wave 
superconductors where the spin-orbit coupling is weak.

\end{abstract}

\begin{multicols}{2}
\narrowtext

Our fascination towards excitations 
carrying quantum numbers distinct from 
electrons in condensed matter systems dated back at least twenty years ago
\cite{Heeger,Robert,PWAnderson,Baskaran}.
It is now widely accepted that there are a variety of
strongly interacting systems
where quasiparticles carry only a fraction
of quantum numbers an electron has.
All known examples, though have been discovered in very diversified
condensed matter systems which bear no similarities at first sight,
appear to share one remarkable common feature.
It is topological excitations interacting with 
electrons one way or the other which make all sorts
of exotic quasiparticles or collective excitations
possible. This also lies in the heart of 
earlier examples discovered in field theories
and mathematical physics\cite{Jackiw,tHooft,Goldstone81}.

In this letter, we scrutinize the
spin ordering and its influences on 
excitations, particularly, on the fractionalization of
quasiparticles and collective spin excitations
in $2D$ spin triplet p-wave superconductors.
Spin triplet superconducting states are believed to exist in
$^{3}He$, many heavy-fermion superconductors and most recently layer 
perovskite $Sr_2 Ru O_4$ 
crystals\cite{Vollhart,Volovik,??,Maeno94}.
We report the existence of a new $2D$ spin triplet p-wave  
superconducting state which is spin rotation invariant. This new state 
is characterized by a finite 
range spin correlation and $hc/4e$ vortices as
the elementary topological excitations. 
The elementary quasiparticles are 
Bogolubov-De Gennes ({\em BdeG}) quasiparticles hosted in 
zero energy skyrmions.  
The spin collective excitations are shown to be fractionalized ones
carrying a half spin and obeying fermionic statistics, by contrast to
the spin wave excitations in spin ordered p-wave
superconducting states({\em SOpSS}s).
We should mention that the general fractionalization pattern in 
some p-wave 
superconductors was recently classified in\cite{Demler}.

For a p-wave superconductor with an order parameter
${\bf d}({\bf k})=\Delta_0 ({ k}_x +i{ k}_y) \exp(i\chi)
{\bf n}$\cite{Vollhart,Volovik},
the Hamiltonian 
in the Nambu space of 
$\Psi=(\psi^+, i\tau_2 \psi)$
can be written as,

\begin{eqnarray}
H=\sigma_3 \epsilon + \sum_{i=x,y} \sigma_{i} \{\partial_i, 
\hat{\Delta}\}_+,
\end{eqnarray}
where $\hat{\Delta}$ is defined as
$\hat{\Delta}=\Delta_0 \exp(i\sigma_3\chi)({\bf n} \cdot {\bf \tau})$ and
$\epsilon({\bf k})=\hbar^2 {\bf k}^2/2 m - \epsilon_F$. 
We use $\sigma$ as the Nambu space Pauli matrix and $\tau$ as
the spin space one.
We assume the spin-orbit interactions are weak and
${\bf n}$ is a unit vector in a sphere $S^2$.
The internal space of the symmetry broken state is ${\cal R}=
[S^1 \times S^2]/Z_2$.
The order parameter observes a discrete
symmetry: $\hat{\Delta}({\bf n},\chi) \rightarrow \hat{\Delta}(-{\bf n}, 
\chi+\pi)$
and represents a quantum spin nematic 
p-wave superconducting state.

{\bf Spin-phase separation}
To obtain an effective theory, we integrate over
the fermionic degrees of freedom and make a gradient expansion.
At low temperatures, we report the result as

\begin{eqnarray}
&& {\cal L}={\cal S}_{ab} {\bf P}_{sa} {\bf P}_{sb} - {\cal T}_0 \Phi^2
+\frac{1}{8\pi}\big( {{\bf E}}^2 +{{\bf B}}^2  \big)
\nonumber \\
&&+{\cal S}^{\alpha\beta}_{ab}\nabla_{a} {\bf n}_\alpha
\nabla_{b} {\bf n}_\beta -{\cal T}^{\alpha\beta} \partial_t
{\bf n}_\alpha \partial_t {\bf n}_\beta
+\frac{\cal N}{4\pi} \epsilon^{\mu\nu\lambda}{\bf A}_\mu {\bf F}_{\nu 
\lambda}.
\label{Lagrangian}
\end{eqnarray}
${\bf P}_{s}=\frac{1}{2}\nabla \chi + e{\bf A}^{em}$ and 
$\Phi=\frac{1}{2}\partial_t \chi +e\phi^{em}$ are the gauge invariant 
momentum and potentials respectively.
The superscript ${\em em}$ is introduced to distinguish the usual 
electric magnetic vector potential ${\bf A}^{em}$ from the
topological field ${\bf A}$ 
defined below in terms of ${\bf n}$. 
${\cal T}_0\nu^{-1}_0$ is a unity at $T=0$
and varies smoothly as a function of the temperature,
$\nu_0$ is the averaged density of states at the fermi
surface. 
${\bf F}_{\mu\nu}=\frac{1}{2}{\bf n} \cdot \partial_\mu {\bf n} \times \partial_{\nu} 
{\bf n}$,
and ${\bf A}$ is the vector potential of ${\bf F}_{\mu\nu}$.
${\cal S}^{\alpha\beta}_{ab}=\delta^{\alpha\beta}{\cal S}_{ab}$,
${\cal S}_{ab}={\rho_0}/2m \tilde{{\cal S}}_{ab}$; and
${\cal T}^{\alpha\beta}=\delta^{\alpha\beta}\nu_0 \tilde{{\cal T}}$.
Finally, $\tilde{\cal S}_{ab}$, $\tilde{\cal T}$ and ${\cal N}$ are
calculated as

\begin{eqnarray}
&& \tilde{\cal S}_{ab}
=\frac{2m}{\rho_0} \int \frac{d^2 k}{(2\pi)^2}
{{\bf v}_a {\bf v}_b}
F({\bf k}),
\tilde{\cal T}=\frac{1}{\nu_0}
\int \frac{d^2k}{(2\pi)^2}
F({\bf k});
\nonumber \\
&& {\cal N}=
\int \frac{d^2k}{4\pi}
\epsilon^{\alpha\beta\gamma}
\frac{{\bf M}_\alpha}{\Delta_0^2}
\frac{{\bf M}_\beta}{\partial {k_x}} 
\frac{{\bf M}_\gamma}
{ \partial {k_y}}
F({\bf k});
\nonumber \\
&& F({\bf k},T)=-
\frac{\partial }{\partial
\epsilon^2({\bf k})}
\frac{2\Delta_0^2(T)}{E({\bf k})}\tan(\frac{E({\bf k})}{2 kT}).
\end{eqnarray}
$\rho_0$, $m$ and $\epsilon_F$ are the density, mass and 
fermi energy respectively; $E({\bf k})=\sqrt{\epsilon^2({\bf k}) 
+\Delta_0^2 {\bf k}^2/k_F^2}$,
and
$\Delta_0$ is the temperature dependent gap.
At zero temperature, ${\cal N}$ is quantized to be unity in 2D.
The director ${\bf M}$ is defined as ${\bf
M}({\bf k})=\big(
v_\Delta {\bf k}_x, v_\Delta{\bf 
k}_y, \epsilon(k)\big)$; $v_\Delta=\Delta_0/k_F$.
When the director ${\bf n}$ is 
a planar vector confined in the $\theta=\pi/2$ equator, i.e.
${\bf n}=(\cos\phi, \sin\phi, 0)$,
the spatial gradient terms coincide with previous results
for the {\em A} phase of $^{3}He$\cite{Anderson,Cross}.
The action is valid 
when the frequency and the wave vector
are smaller than $\Delta_0(T)$ and $\xi^{-1}_0(T)=\Delta_0(T)/v_F$
respectively.

Eq.\ref{Lagrangian} suggests a few important properties of 
the quantum spin
nematic p-wave superconductors. First of all, the dynamics of
spin ${\bf n}$ and phase $\chi$ is completely decoupled at 
the low frequency limit(except the entanglement due to a $Z_2$ projection 
in the functional integral\cite{Demler}, which we will
not discuss in this paper). 
It reflects spin-phase
separation in a p-wave superconductor.
Second, 
for an isotropic fermi surface which interests us in this article, ${\cal 
S}_{ab}=\delta_{ab} \rho_s(T)/2m$,
and $\rho_s(T)$ is the temperature dependent superfluid density which
vanishes at the critical temperature $T_c$.
So the spin and phase dynamics 
are characterized by an $O(3)$ $\sigma$-model (NL$\sigma$M) and  
an $xy$ model respectively. 
At the mean field approximation,
${\bf n}={\bf e}_z$ and $\chi$ is a constant. This corresponds to
a conventional {\em SOpSS}. There are three
Goldstone modes; two of them are spin waves $\delta {\bf n}
=(1, \pm i, 0)$ with a linear dispersion.
In an isotropic case, $\tilde{\cal S}_{ab}=\delta_{ab}\tilde{ {\cal S}}$;
the spin wave velocity is
$v_{s}(T)=v_F \sqrt{{\tilde{\cal S}}/{2\pi \tilde{\cal 
T}}}$. And the last mode is the usual plasma wave, with a dispersion
$\omega=\sqrt{2\pi e^2\rho_0 k/m}$ in $2D$ at $T=0$.

The topological term was previously derived in\cite{Volovik89,Volovik90}. 
This term, orignating from the broken time reversal and parity
symmetries, determines 
the topological order
in the fields $F_{\mu\nu}$ and defines the structure of quasiparticles.
Implications of topological terms in other unconventional superconductors 
were also explored recently\cite{Senthil98,Goryo99,Read00}. Here,
we investigate the spin ordering,
zero energy spin textures and quasiparticles  
based on Eq.\ref{Lagrangian}.
Let us emphasis that Eq.\ref{Lagrangian} is valid as far
as the quasiparticles are gapped and the gradient expansion 
is possible; physically, it says all low lying collective excitations 
below the BCS energy gap are correctly described by the action.

{\em {\bf Spin ordering}}
Because of an extra branch of Goldstone modes in the
spin sector, the spin order is more fragile than the phase order
in the problem. In $2D$, this provides a unique possibility of spin 
disordered 
p-wave superconducting states ({\em SDpSS}s) where the $S^2$-symmetry is 
restored and only the $U(1)$-symmetry is broken. 
Such a state which is rotation invariant in nature differs from 
the conventional
{\em SOpSS}s where the $S^2$ symmetry is broken and
there is a long range order in ${\bf n}$.

The finite temperature phase diagrams of the $O(3)$ Nonlinear $\sigma$ 
Model (NL$\sigma$M) were 
previously analyzed in great details\cite{CNH}.
In the current situation, 
just as the superfluid velocity $\rho_s(T)$,  
all coefficients in the action, ${\cal S}^{\alpha\beta}_{ab},
{\cal T}_{ab}, {\cal S}_{ab}, {\cal T}$  
depend on temperatures
because of quasiparticle excitations. Taking this into account, we 
arrive at the following results in $2D$.


When $\Delta_0 \ll \epsilon_F$, the spin order is
established at zero temperature and the correlation length 

\begin{equation}
\xi_2=\frac{v_s(T)}{ \Delta_s(T)},
\Delta_s=T \exp \big( -\frac{2\pi [\rho_s(T)/2m 
-\Gamma]}{T} \big)
\label{order}
\end{equation}
is finite only at finite temperatures(in a saddle point approximation). 
Here $\Gamma \sim \Delta_0(T)$.
For most of p-wave superconductors,
the gap energy is about $1K$ and 
the 
Fermi energy of order $1eV$, 
the superconductors are spin ordered at zero 
temperature.
However, the
liquids are spin disordered (in the 
absence of spin-orbit coupling)
at any finite temperature as shown in Eq.\ref{order}.
On the other hand, when $\Delta_0$ is much larger than the fermi energy, 
the spin long range order could be spoiled by quantum fluctuations
and the rotation invariance is preserved\cite{MF}.
We will be concerned with both situations, {\em thermal} and 
{\em quantum} $2D$ {\em SDpSS}s in the following 
discussion.

{\em {\bf Zero energy skyrmions} }
One of the most important feature of the {\em quantum} {\em SDpSS} 
is the existence of topological order and consequently 
topological stable zero energy skyrmions in the absence of
spin stiffness.
In $(2+1)$ space ${\bf x}=(\tau, {\bf r})$, it is convenient to introduce
a field,
${\bf H}_{\eta}=\frac{1}{2}\epsilon^{\eta\mu\nu}{\bf F}_{\mu\nu}$.
${\bf H}_\tau={\bf F}_{xy}$ represents $U(1)$ magnetic fields
along $z$ direction, ${\bf H}_x={\bf F}_{y\tau}$ and
${\bf H}_y={\bf F}_{\tau x}$ are the $x,y$-components of the electric
field.
To facilitate a calculation at finite temperatures, 
the perimeter along $\tau$-direction $L_\tau$ is taken
to be finite, i.e. $L_\tau=(kT)^{-1}$.
Consider a rotating skyrmion terminated at the origin in
a $S^2 \times S^1$ space
${\bf n}(\rho, \phi)=\big( \sin\theta (\rho) \cos(\tilde{\phi}),
\sin\theta(\rho)\sin(\tilde{\phi}), \cos\theta(\rho) \big)$
where

\begin{eqnarray}
&& \tilde{\phi}=Q_m \phi-\gamma(\tau), \nonumber \\
&& \theta(\rho,\tau)=2 arcos \frac{\rho}{\sqrt{\rho^2 +v_s^2\tau^2}}
\Theta(\tau),
\nonumber \\
&& \gamma(\tau+ L_\tau) -\gamma(\tau)=N 2\pi.
\label{monopole}
\end{eqnarray}
One can confirm that
$\nabla \cdot {\bf H} = Q_m 2\pi \delta(\tau)\delta({\bf r})$,
corresponding to a space-time monopole of charge $Q_m$ in $2+1$d.
As $\rho, \tau$ approach infinity, ${\bf H}(\rho,\tau)$
becomes vanishly small.
The action of this Euclidean space monopole event
is finite ($a \sim 1$),

\begin{equation}
{\cal S}_{m.}= \frac{a\Delta_0}{16\pi \Delta_s }+i\gamma_B,
\gamma_B=\frac{Q_m{\cal N}}{4} [\gamma (L_\tau)-
\gamma(\tau_0)].
\end{equation}
However, it has a Berry's phase due to the topological term, which 
characterizes a rotation of the skyrmion during its duration.
$\gamma_B(0)$ obviously depends on the temporal coordinate at which the
skyrmion is terminated, leading to destructive interferences between 
monopoles centered at different $\tau_0$
with different rotation angles $\gamma_B$.

As a result,
the fluctuations of space-time monopole events per unit volumn
are($c\sim 1$)

\begin{equation}
< Q_m^2>
=\delta({\cal N})
\frac{\Delta_0}{c \xi_0^2} \exp(-\frac{a \Delta_0}{16\pi \Delta_s}). 
\label{Rate}
\end{equation}
Eq.\ref{Rate} shows that at any finite ${\cal N}$ all monopole 
events are suppressed 
due to destructive
interferences.
It also implies that for ${\cal N}\neq 0$
the ground state has an infinite-fold degeneracy
compared with that of ${\cal N}=0$.

There are at least two important intraconnected consequencies of
the destructive interferences.
First, Eq.\ref{Rate} indicates
the conservation of the Skyrmion charges at ${\cal N}\neq 0$
in a {\em quantum} {\em SDpSS}, that is in the absence of the spin 
rigidity.
If we define
$c_{w}\big(\{ {\bf n}({\bf r}) \}\big)=\frac{1}{2\pi}\int dx dy {\bf
H}_z$
as the total number of Skyrmions living on the 2D sheet, 
in the presence of space-time monopoles  $Q_m$ at $\{ {\bf r}^m,\tau_m \}$,
\begin{equation}
\frac{\partial c_{w}(\tau)}{\partial \tau}=\sum Q_m
\delta(\tau-\tau_m).
\label{rate}
\end{equation}
A space-time monopole
essentially connects a trivial vacuum to a Skyrmion configuration
and causes a change in the topological charge $c_{w}$ by one unit.
At ${\cal N}=1$, following Eqs.\ref{rate}, \ref{Rate}, we conclude 
that a skyrmion 
whose energy could vanish in the absence of the spin stiffness,
is a well-defined topological configuration in a {\em quantum} 
{\em SDpSS}. This remarkable feature which doesn't exist at 
${\cal N}=0$ is also a consequency of 
a zero energy fermionic
mode hosted by instantons.

Second, the suppression of monopole
events leads to very distinct behaviors of fields ${\bf F}_{\mu\nu}$
in {\em SDpSS}s.
The Wilson-loop integral defined as
${\cal W}_{U(1)}=\big<{\cal P}\exp\big( i\oint {\bf A} \cdot d{\bf r}
\big)\big>$ has
different asymptotical behaviors in the large loop limit in the  
presence or absence of topological
order in $c_{w}$. 
When the topological charge $c_{w}$ is conserved
at any finite ${\cal N}$, 
${\cal W}_{U(1)}=\exp(- L_c C_1)$ ($L_c$ is the perimeter of the Wilson
loop) and the gauge fields are deconfining.
This is true for a {\em quantum SDpSS} at zero and finite temperatures as 
far as 
$L_\tau$ is longer than the
duration of space-time monopoles.
However, in a {\em thermal} {\em SDpSS}, $c_{w}$ is unconserved and
the gauge fields are confining
(except around the quantum critical point which I will not discuss here).

{\em {\bf Quasi-particles}}
We now employ the generalized Bogolubov-De Gennes
equation to study 
the properties of quasiparticles in {\em SDpSS}s.
In the presence of a topological configuration of
${\bf n}({\bf r})$, it is convenient to
introduce a gauge transformation
$\Psi\rightarrow U_s({\bf n}) U_c(\chi) \Psi$ and work in a rotated 
representation; then 
one obtains a new Hamiltonian 

\begin{eqnarray}
H=\sigma_3 \epsilon(
i\hat{\nabla}) 
+ v_\Delta \sum_{i=1,2} 
\{ \sigma_{i}\tau_3,
i\hat{\nabla}_i \}_+.
\label{Hamiltonian}
\end{eqnarray}
Here $v_\Delta =\Delta_0/k_F$,
$i\hat{\nabla}=i\nabla-{\bf A}_{c} - {\bf A}_{s}$ is a covariant 
derivative.
We have defined 
$U_s^{-1}{\bf d}\cdot {\bf \tau} U_s=\tau_3$,
$U_c^{-1}\sigma_i\exp(i\sigma_3 \chi) U_c=\sigma_i$.
The vector potentials are defined 
in terms of the $U(1)$ rotation $U_c$  and $SU(2)$ rotation
$U_s$ as
${\bf A}_{c\mu}=i U_c^{-1}\partial_\mu U_c=\sigma_3 ({\bf A}^{em}_\mu 
+\frac{1}{2}\partial_\mu \chi)$,
${\bf A}_{s\mu}=i U_s^{-1}\partial_\mu U_s={\bf \tau}_\alpha$ $\cdot 
{\bf W}_\mu^\alpha$ ($\mu=0, 1,2$,
stands for coordinates in $1+2$ dimension space.).
An explicit calculation also shows that ${\bf W}^3_\mu={\bf A}_\mu$.
At last, the corresponding Lagrangian density is

\begin{eqnarray}
{\cal L}_{BdeG}=
\Psi^+ \big (
\hat{\partial_\tau}
-{\cal H}({\bf A}^{em}_\mu, {\bf A}_{s\mu})
\big) \Psi,
\end{eqnarray}
and $\hat{\partial_\tau}=
\partial_\tau-\sigma_3 A^{em}_0 - {\bf \tau}\cdot {\bf 
A}_{s0}$.

Following Eq.\ref{Hamiltonian},
besides a usual electric magnetic charge defined with
respect to ${\bf A}_{e.m.}$ fields, a {\em BdeG} 
quasiparticle
also carries a unit $U(1)$-charge with respect to ${\bf A}_\mu$ fields
and is minimally coupled with ${\bf F}_{\mu\nu}$.
The energy of a {\em BdeG} particle is determined by 
the Wilson loop integral of ${\bf A}$.
In {\em quantum} {\em SDpSS}s, 
the Wilson loop integral decays exponentially as a function
of the perimeter of the loop.
The interactions between {\em BdeG} quasiparticles
mediated by the topological fields ${\bf F}_{\mu\nu}$ are
rather weak and
the {\em BdeG} quasiparticle energy is finite.
But most importantly,
in this case, skyrmions themselves carry $U(1)$ charges with respect to
the fields ${\bf A}_\mu$.
This is indicated in Eq.\ref{Lagrangian}
if we introduce the skyrmion density-current density as
$4\pi {\bf j}_\mu={\cal N}\epsilon_{\mu\nu\eta}\partial_\nu {\bf A}_\eta$
and express the topological term in a form of minimal
coupling ${\bf j}_\mu {\bf A}_\mu$.
By minimizing the
action of ${\cal L} +{\cal L}_{BdeG}$ with respect to ${\bf A}_0$,
${\bf A}^{em}$
and taking ${\cal N}=1$ at low temperature limit, we
do obtain a saddle point equation
$4 \pi <\Psi^+\tau_3\Psi>={\bf e}_z \nabla \times {\bf A}$,
$<\Psi^+\sigma_3\Psi>=0$.  This indicates
that a skyrmion configuration carries
a half spin but no charge.
In other 
words, a spin $\frac{1}{2}$ but chargeless 
{\em BdeG} quasiparticle is hosted by, or 
confined with a 
skyrmion, with the confinement mediated by the spin fluctuations.

To examine the {\em BdeG} quasiparticles dressed with spin textures,
we consider a skyrmion in polar coordinates $(\rho,\phi)$. The
director has a spatial distribution as ${\bf n}(\rho,\phi)
=(\sin\theta(\rho)\sin\phi, \sin\theta(\rho)\cos\phi, \cos\theta(\rho))$;
$\theta(\rho)$ is a smooth function of $\rho$, the asymptotics
of which is 
$\theta(\rho=0)=0$ and $\theta(\rho\rightarrow \infty)=\pi$. 
The corresponding ${\bf A}$ field can be chosen as

\begin{equation}
{\bf A}=\frac{1-\cos\theta(\rho)}{2 \rho\sin\theta} {\bf e}_\phi, \nabla \times 
{\bf 
A}=\frac{\sin\theta(\rho)}{2 \rho}\frac{\partial \theta(\rho)}{\partial 
\rho} {\bf e}_z.
\end{equation}
The $SU(2)$ field ${\bf W}^\alpha$ at $\rho\rightarrow \infty$ 
can be shown to take a simple form;
${\bf W}^3_i={\bf A}_i$($i=1,2$), ${\bf W}^3_0=0$ and ${\bf W}^1_\mu={\bf 
W}^2_\mu=0$.

A {\em BdeG} quasiparticle remains gapped in a texture.
However, following Eq.11 when a spin-$1/2$ {\em BdeG} particle 
moves in a closed circle of radius $\rho$ in a skyrmion
defect, 
it acquires a Berry's phase of $\pi
[1-\cos\theta({\rho})]$ which approaches $2\pi$ at
an infinity $\rho$.  
Consequently, under the interchange of coordinates, the two-body 
wave function
of composite quasiparticles
acquires an additional $\pi$ phase because of hosting skyrmions,
and $\Psi({\bf r}_1, {\bf r}_2)=\Psi({\bf r}_2, {\bf r}_1)$,
which also follows the linking number theorem for
skyrmions\cite{Wilczek}.
These composite quasi-particles are therefore Bosons.
We also observe the {\em BdeG} 
quasiparticles are charge 
neutral 
at $\epsilon({\bf k})=0$ 
with respect to
an {\em em} field; they also carry zero $U(1)$ 
charges so to minimize the interaction between composite 
excitations.
The life time of the quasiparticles is limited by the life time
of zero energy skyrmions;
for the {\em quantum} disordered case,
the zero energy skyrmions are stable even at low temperatures.

In {\em thermal} {\em SDpSS}s, the suppression of space-time monopoles
is incomplete since $L_\tau$ is longer than the monopoles' duration 
$\Delta_s^{-1}$.
The gauge field then is confining and the {\em BdeG} quasi-particles
form bound states, with zero or one total spin.
This unexpected feature should be observed in future experiments 
on some layered p-wave superconductors.





{\bf Collective spin excitations}
The nature of the collective spin excitations
in an {\em SDpSS} can be 
explored in a spinor representation of Eq.\ref{Lagrangian}.
By introducing $\eta^+ {\bf \tau} \eta ={\bf n}$, $\eta=(
\eta_1, \eta_2)^T$ and $\eta^+ \eta=1$, we obtain for $\eta$
the following Lagrangian in {\em SDpSS}s,

\begin{equation}
{\cal L}_{\eta}=\frac{1}{2f^2}|(i\partial_\mu -{\bf A}_\mu) \eta|^2
+\frac{\Delta_s(T)}{\Delta_0(T)} \eta^+\eta + 
\frac{\cal N}{4\pi} \epsilon^{\mu\nu\lambda}{\bf A}_\mu {\bf F}_{\nu
\lambda}.
\label{eta}
\end{equation} 
And $\eta$ is a bosonic field carrying a unit charge with respect to
${\bf A}$ fields and spin $1/2$.
In Eq.\ref{eta}, $2f^2=2 m\Delta_0 /
\sqrt{\tilde{\cal S}\tilde{\cal T}}\rho_0$;
we have introduced the following rescaling:
$t\rightarrow t \xi_0/v_s$, ${\bf r} \rightarrow {\bf r} \xi_0$.

In quantum {\em SDpSS}s,
an $\eta$- 
quantum is bound with a skyrmion such that the bound state becomes a 
fermion\cite{Dzyaloshinskii}.
Each spin one spin wave excitation which is an elementary excitation
in an {\em SOpSS}, is fractionalized into two elementary 
fermionic spinors hosted in skyrmions in quantum {\em SDpSS}s.
Each spinor-skyrmion composite is a spin-$1/2$ excitation 
carrying no $U(1)$ charge, by contrast to a bare $\eta$ excitation.
In the thermal {\em SDpSS}s , the spin collective excitations 
are spin-wave ones with spin one.





{\em {\bf $\frac{\hbar c}{4 e}$ vortices}}
For the sake of completeness, I am also listing some
properties of vortices.
The linear defects in a symmetry broken state
with an internal space ${\cal R}=[S^1\times S^2]/Z_2$ have been recently 
discussed extensively
in the context of Bose-Einstein condensates of 
$^{23}Na$\cite{Zhou1}.
In {\em SOpSS}s, the linear defects are superpositions
of $h c/4e$ vortices and $\pi$-disclinations because of the $Z_2$
symmetries in the problem.
And a bare $hc/4e$ vortex is forbidden because of the catastrophe
of a cut. In {\em SDpSS}s, however, $hc/4e$ vortices can exist by their
own right and are elementary excitations.

The {\em SDpSS} discussed
here has the following order parameters:
$<\hat{ \Delta}>=0 , Tr< \hat{\Delta}\hat{ \Delta} > \neq 0, < \exp(i\chi) 
>\neq 0$.
The existence of $SC^*$ with 
$<\hat{ \Delta}>=0$ , $Tr<\hat{ \Delta}\hat{ \Delta} > \neq 0$, $< 
\exp(i\chi) >= 0$, and other fractionalized states
examined in \cite{Demler}
appears to be beyond the model studied here. 
Physically, the {\em SDpSS} has
Josephson oscillations of $2eV$ frequency while in $SC^*$ the
frequency is $4eV$.

In the presence of spin-orbital couplings, 
the mean field solution indicates that the
director of ${\bf n}$ points along $\pm {\bf e}_z$ direction and
the internal space is $[Z_2 \times S^1]/Z_2$.
However, at an energy scale higher than the spin-orbit coupling ones, 
${\bf n}$ would be free to rotate on a two-sphere.
The spin order-disorder transition still takes place at a finite 
temperature
below the superconductor-metal transition temperature $T_c$
when the spin-orbit scattering rate is much smaller than $\Delta_0 (0)$.
As the spin is disordered, 
the above discussions
on the spin textures and {\em BdeG} quasiparticles are still valid.

In conclusion,
we also would like to remark that some aspects of 
the {\em BdeG} quasiparticles
in spin disordered superconductors
considered here 
reminisce the chiral-bag defect model for the nucleon
\cite{Skyrme,Goldstone}.
The presence of spin-$1/2$ bosonic chargeless {\em BdeG} excitations
in a quantum {\em SDpSS}
is an example of fermi number 
fractionalization; it belongs
to the same class phenomenon as the 
mid-gap quasiparticles hosted in
domain wall excitations in one dimension polyacetylene\cite{Heeger}
and the statistical transmutation proposed in some magnetic models
\cite{Dzyaloshinskii,Wiegmann,Kivelson,Wen}.
It is my pleasure to thank P. W. Anderson, E. Demler, D.Khmelnitskii,
K. Schoutens and F. Wilczek for useful discussions, 
and X. G. Wen for a conversation on instantonic zero modes.
Finally, I am grateful to
P.Wiegmann for explaining to me the global anomalies
and patiently scrutinizing my arguments.

\end{multicols}


\begin{thebibliography}{99}
\bibitem{MF}This is supported by a saddle point calculation.
\end{thebibliography}
\end{document}